\documentclass[aps,prl,superscriptaddress,preprintnumbers,reprint]{revtex4-1}

\newcommand{\bea}{\begin{eqnarray}}
\newcommand{\eea}{\end{eqnarray}}
\newcommand{\be}{\begin{equation}}
\newcommand{\ee}{\end{equation}}
\newcommand{\mw}   {\mbox{$m_{\scriptsize W}$}}

\newcommand{\mwzero}   {\mbox{$m_{\scriptsize W,0}$}}

\newcommand{\seffl}{\sin^2\theta_{\mathrm{eff}}^{\ell}\,}
\newcommand{\ptlep}{p_\perp^\ell}
\newcommand{\ptz}{p_\perp^Z}

\newcommand{\nnpdf}{{\tt NNPDF30\_nlo\_as\_0118\_1000}\,}
\newcommand{\powheg}{{\tt POWHEG-BOX}\,}
\newcommand{\pythia}{{\tt Pythia8.2}\,}

\usepackage{graphicx}
\usepackage{xcolor}
\usepackage[normalem]{ulem}

\begin{document}

\preprint{CERN-TH-2019-139, DESY 19-169, PSI-PR-19-21}
\title{Parton Density Uncertainties and the Determination of Electroweak Parameters at Hadron Colliders}
\author{Emanuele Bagnaschi}
\email[]{emanuele.bagnaschi@psi.ch}
\affiliation{Paul Scherrer Institut, CH-5232 Villigen, Switzerland}
\author{Alessandro Vicini}
\email[]{alessandro.vicini@mi.infn.it}
\affiliation{TH Department, CERN 1 Esplanade des Particules, Geneva 23, CH-1211, Switzerland and\\
  Dipartimento di Fisica ``Aldo Pontremoli'', University of Milano and INFN Sezione di Milano,
  Via Celoria 16, 20133 Milano, Italy}

\date{\today}

\begin{abstract}
  We discuss the determination of electroweak parameters from hadron
  collider observables, focusing on the $W$ boson mass measurement.
  We revise the procedures adopted in the literature to include in the experimental analysis
  the uncertainty due to our imperfect knowledge of the proton structure.
  We show how the treatment of the proton parton density functions (PDFs) uncertainty
  as a source of systematic error, leads to the automatic inclusion in the fit
  of the bin-bin correlation of the kinematic distributions with respect to PDF variations.
  In the case of the determination of $M_W$ from the charged lepton transverse momentum
  distribution, we observe that the inclusion of this correlation factor
  yields a strong reduction  of the PDF uncertainty,
  given a sufficiently good control over all the other error sources.
  This improvement depends on a systematic accounting of the features of the QCD-based PDF model,
  and it is achieved relying only on the information available in current PDF sets.
  While a realistic quantitative estimate requires to take into account
  the details of the experimental systematics,
  we argue that, in perspective, the proton PDF uncertainty
  will not be a bottleneck for precision measurements.
\end{abstract}

\maketitle

\section{Introduction}

The values of the $W$ boson mass $\mw$ and of the sinus of the
leptonic effective weak mixing angle $\seffl$ are very precise
predictions in the electroweak (EW) sector of the Standard Model (SM)
and allow stringent tests at the level of the quantum corrections.
The measurements of these two parameters
at the Tevatron and at LHC
indicate \cite{Aaltonen:2013iut,Aaboud:2017svj,Aaltonen:2018dxj,ATLAS:2018gqq,Sirunyan:2018swq,Aaij:2015lka}
the imperfect knowledge of the proton structure
as one of the main sources of systematic uncertainty of theoretical origin.
The latter affects the computation of the templates used in the fit of
the kinematic distributions and eventually the determination of the EW
parameters.

The proton collinear parton distribution functions (PDFs) suffer from
different uncertainties of experimental as well as theoretical origin.
The impact of the error of the data from which the PDFs are extracted
is represented by sets of functions,
Hessian eigenvectors or Monte Carlo replicas,
that span in a statistically significant way
the functional space of all possible parameterisations.
The propagation of the experimental error in the prediction of any observable
is achieved by simply repeating the evaluation of the latter several times,
with all the available members of the PDF set; mean and standard
deviation are eventually computed, the latter being the propagation
of the experimental error to the observable under study.
All the members of a PDF set share some common theoretical features,
like the fact that they all obey the perturbative QCD (PQCD) evolution equations
and sum rules, and are thus correlated with each other.
While this correlation is automatically included in the propagation of the
experimental PDF error to the prediction of any observable,
the determination of a parameter extracted from the simultaneous fit of several
observables requires a careful discussion.

In the case of the $W$ boson mass determination,
the role of the PDFs has been discussed in the articles presenting the
experimental results, for those PDF sets used in the analyses, while a
more general comparison of different parameterisations has been
presented in Refs.~\cite{Bozzi:2011ww,Bozzi:2015hha};
in all cases the common outcome is that an uncertainty $\Delta^{\mathrm{PDF}}\mw$
at the $10$~($20$)~MeV level is expected in the lepton-pair transverse mass
(lepton transverse momentum) case, with the precise value depending on several details of the
analyses and on which parameterisations are included in the study.
All these studies considered the fit of a kinematic distribution,
by combining the information of different bins weighed by their statistical and systematic errors,
but neglecting any bin-bin correlation with respect to PDF variations.
In Ref.~\cite{Bozzi:2015zja} the dependence of the uncertainty on the rapidity
range included in the acceptance region was exploited to quantify the
benefit given by an $\mw$ measurement at LHCb to the final combination
of all the available results, in terms of a reduced PDF uncertainty.
The possibility of a systematic extraction of very precise information
about the Drell-Yan parton-parton luminosities has been studied in
Refs.~\cite{Manca:2017xym,Bagnaschi:2018dnh,Bacchetta:2018lna,Farry:2019rfg,Bianchini:2019iey}, aiming at a better modelling of the
initial state and to a consequent reduction of the PDF error on $\mw$.
The impact of measurements at different colliders and energies has been
scrutinised in Ref.~\cite{ATL-PHYS-PUB-2018-026}.

We plan to revise the propagation of the PDF
uncertainties of experimental origin, in the determination of a
parameter obtained via the fit of a kinematic observable.

\section{Bin-bin correlation and template fit definitions}
At hadron colliders $\mw$ is determined in the charged-current
(CC) Drell-Yan (DY) process, from the measurement of observables
such as the charged-lepton transverse momentum
$d\sigma/dp_\perp^{\ell}$ and the lepton-pair transverse mass
distributions, for which the Jacobian peak enhances the
sensitivity to the position of the pole of the $W$ propagator.
The finite rapidity detector acceptance and other kinematical constraints
induce a sensitivity of the shape of these observables,
defined in the transverse plane, to the initial state proton collinear PDFs.

\begin{figure}[!h]
  \includegraphics[width=80mm,angle=0]{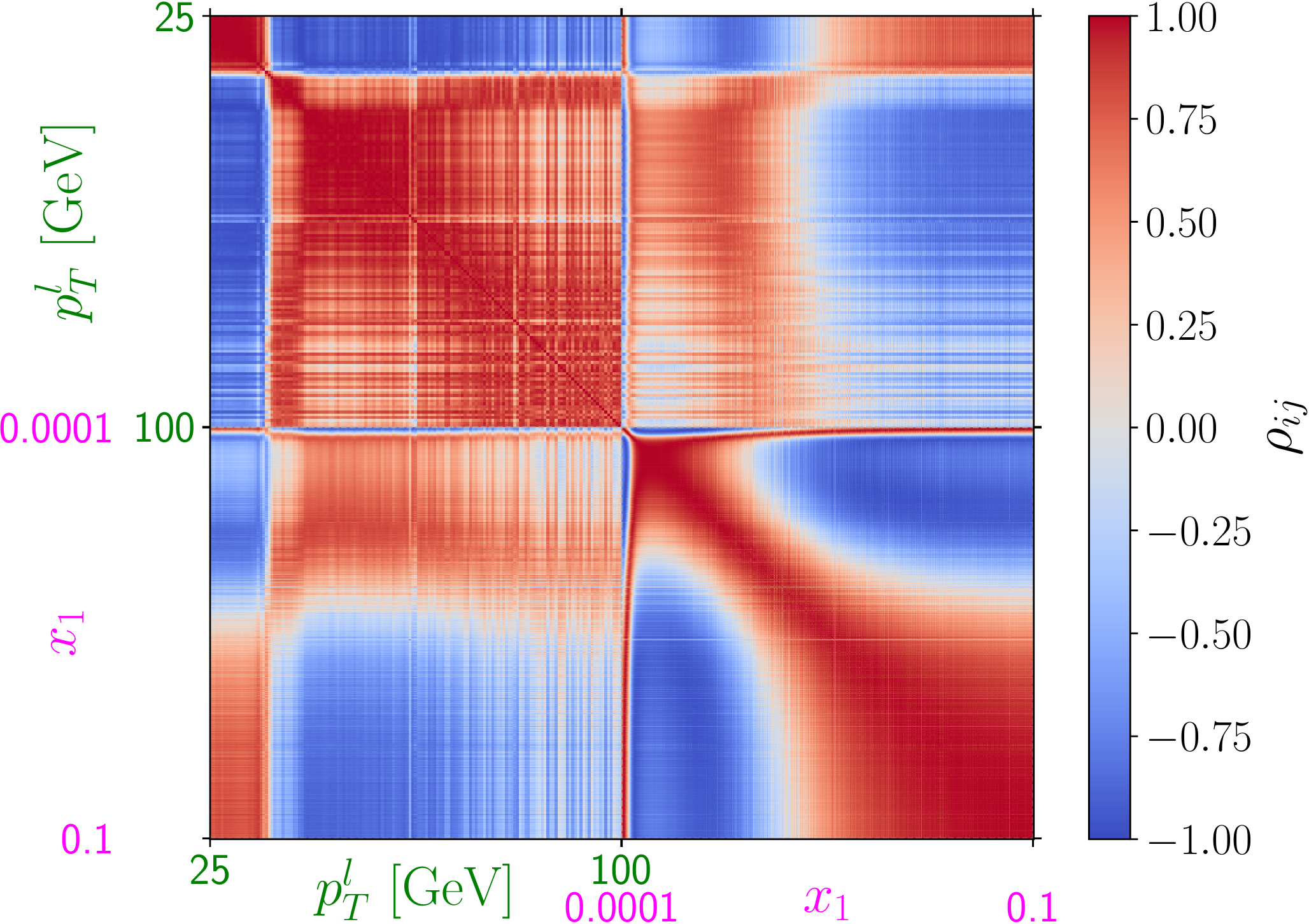}
  \caption{\label{fig:corr-pt-x-shape}
    Correlation with respect to PDF replica variations of the differential cross sections
    with respect to the charged lepton transverse momentum and to the partonic $x_1$ variable.
    Results obtained with normalised distributions.
    We show the self correlation of $d\sigma/dx_1$ (lower right),
    of $d\sigma/dp_\perp^{\ell}$ (upper left) and the cross correlation of the two distributions.
Fluctuations due to finite MC statistics are visible as a stripe-like pattern in the plots.
  }
\end{figure}

The PDF uncertainty, represented by a set of replicas with $N_{rep}$ members, affects
the normalisation but also the shape of the observables.  Different
bins of the same distribution are correlated with respect to a PDF
replica variation, as it can be seen in
Fig.~\ref{fig:corr-pt-x-shape},
because of kinematic constraints and due to the theoretical
framework shared by all the replicas.
In Fig.~\ref{fig:corr-pt-x-shape}
the sudden and strong change of sign of the
$d\sigma/dp_\perp^\ell$ self-correlation is quite evident across the Jacobian peak at
$p_\perp^\ell \sim 40$ GeV; the self-correlation of the $d\sigma/dx_1$
distribution also signals the existence of two partonic $x$ ranges,
below and above $x\sim 4\cdot 10^{-3}$. The cross correlations
thus establish a link between the parton-parton luminosities, i.e.~the
source of the PDF uncertainty, and the $d\sigma/dp_\perp^\ell$
distribution, from which $\mw$ is determined, with a non-trivial
underlying correlation pattern.

The determination of a Lagrangian parameter from a kinematic
distribution via a template fit requires the choice of a Lagrangian density
(in our case the SM one) and of a tool that simulates the observables
computed in that model
in a well defined setup.
The simulation tool is fully specified by
the choice of a proton PDF parameterisation,
while the parameter, e.g.~$\mw$,
is left free to vary when comparing to the experimental data.
In this construction the PDF replicas represent
a one parameter family of models to analyse the data.

The equivalence of the replicas in the proton description represents a source of theoretical systematic
error, when we try to determine $\mw$ from the fit
of a kinematic distribution.
We account for this systematics in the following $\chi^2$  definition:
\be
\chi^2_{k} =\!\!\!\!
\sum_{i\in bins}\!\!\!
\frac{\left(
  ({\cal T}_{0,k})_i-({\cal D}^{exp})_i-\sum_{r\in {\cal R}} \alpha_r  ({\cal S}_{r,k})_i
  \right)^2}{\sigma_i^2}~+~
\sum_{r\in {\cal R}} \alpha_r^2
\label{eq:chi2alphas}
\ee
where, in the bin $i$ of the distribution, we have the following quantities:
${\cal T}_{0,k}$ is our fitting model based on the average replica $0$ of the PDF set ${\cal R}$
and it has been computed with the $k$-th $W$-boson mass hypothesis $m_{\scriptsize W,k}$;
${\cal D}^{exp}$ is the experimental value and $\sigma_i^2$ is its statistical error;
the differences ${\cal S}_{r,k}={\cal T}_{r,k}-{\cal T}_{0,k}$ are computed
for each member $r$ of the PDF set and are treated as nuisances
with fit parameters $\alpha_r$. The quadratic penalty factor $\sum_{r\in {\cal R}} \alpha_r^2$ corresponds
to having assumed a Gaussian penalty for the replicas with respect to the central replica of the set.
Since the templates are in general affected by statistical Monte Carlo  and experimental errors,
we should take that into account by considering $\sigma_i^2 = (\sigma_i^{exp,stat})^2 +(\sigma_i^{MC})^2$.

By repeating the minimisation of $\chi_{k}^2$, with respect to the $\alpha_r$,
for different values of $m_{\scriptsize W,k}$,
the minimum of the sequence labelled by $k$ selects the preferred $\mw$ value and
the $\Delta\chi^2=1,4,9$ rules identify the 1,2,3 standard deviations intervals due to the PDF uncertainty.
For a given $m_{\scriptsize W,k}$, the minimum of the $\chi^2$ expression in Eq.~\ref{eq:chi2alphas}
can be written \cite{CDFnote} with the bin-bin covariance matrix
computed with respect to PDF variations
and including the statistical and systematic error contributions
\footnote{
A similar approach has been adopted in
Ref.~\cite{CMS:2014mna} for the $\alpha_s$ measurement.  }.
\bea
&&
\chi^2_{k,min}\!\!=\!\!\!\!\!\!\!\!\!
\sum_{(r,s)\in bins}\!\!\!\!
({\cal T}_{0,k}-{\cal D}^{exp} )_r
\left( C^{-1} \right)_{rs}\!
({\cal T}_{0,k}-{\cal D}^{exp} )_s
\label{eq:chi2cov}\\
&&
C=\Sigma_{PDF}+\Sigma_{stat}+\Sigma_{MC}+\Sigma_{exp,syst}
\label{eq:covcomponents}\\
&&
(\Sigma_{PDF})_{rs} \!\!=\!\! \nonumber \\
&&\qquad\quad
\langle
({\cal T}-\langle {\cal T} \rangle_{\small PDF})_r
({\cal T}- \langle {\cal T} \rangle_{\small PDF}  )_s
\rangle_{\small PDF}
\label{eq:sigmacov}\\
&&
\langle {\cal O} \rangle_{\small PDF} \equiv \frac{1}{N_{cov}}\sum_{l=1}^{N_{cov}} {\cal O}^{(l)}
\label{eq:pdfaverage}
\eea
where $\Sigma_{stat}$ is a diagonal matrix with the statistical variances on each bin of the distribution,
estimated for a given integrated luminosity ${\cal L}$, $\Sigma_{MC}$ is the diagonal matrix of the squared
Monte Carlo error of the templates and $N_{cov}$ is the number of PDF replicas
used to compute the PDF covariance matrix \footnote{
In the external expectation value we divide by $N_{cov}-1$.
  }.
  We introduce in the full covariance matrix an additional term $\Sigma_{exp,syst}$ to account for experimental systematics,
  although their faithful description
  depends  on the details of each experiment.
  In Eq.~\ref{eq:covcomponents}
  we approximate $\Sigma_{exp,syst}$ by using the detector model of the Compact Muon Solenoid (CMS) presented in Ref.~\cite{Manca:2016fpw}.
We stress that in this note all the replicas are treated as equivalent,
i.e. we do not anticipate the impact that future measurements may have in reducing the PDF uncertainty.
  The approach that we are proposing to  include the PDF uncertainty on an EW parameter
has to be compared with what has been used in the past, e.g.~in Refs.~\cite{Bozzi:2011ww,Bozzi:2015hha},
where
the analysis relied on the minimisation of a $\chi^2$ defined as
\be
\chi^2_{k,r,\mathrm{no-cov}}=
\sum_{i\in bins} ({\cal T}_{0,k}-{\cal D}_r )_i^2/\sigma_{i}^2
\label{eq:chi2nocov}
\ee
treating the contributions of different bins as independent
and weighing them with their statistical error;
the templates were generated with the central PDF replica 0, for different mass hypotheses $k$;
the distributions, computed with $N_{rep}$ different replicas, were treated as independent pseudodata
and the minimisation was repeated separately for each of them;
the resulting $N_{rep}$ preferred $m_{{\scriptsize W},r}$ values were eventually analysed
computing mean value and standard deviation and ignoring the associated values of $\chi^2_{k,r,\mathrm{no-cov}}$;
the standard deviation was taken as the estimate of the PDF uncertainty.
A similar $\chi^2$ definition, including only diagonal contributions,
has been used up to now by the experimental collaborations  at Tevatron and LHC.

\section{Numerical results}

We perform all the simulations using the CC-DY event generator
provided in the \powheg~\cite{Barze:2012tt,Barze:2013fru},
showered with \pythia~\cite{Sjostrand:2014zea},
setting $\sqrt{S}=13$~TeV. We restrict ourselves to $W^{+}$ production without
hindering the generality of our arguments. We apply the acceptance cut $| \eta_l | < 2.5$.
We use for our analysis the PDF set \nnpdf \cite{Ball:2014uwa}, featuring $N_{rep}=1000$ replicas.

In Eq.~\ref{eq:chi2cov},
the templates are computed using the replica 0 of the PDF set,
scanning $\mw$ with a 1 MeV spacing
in the interval $\mw\in [80.035, 80.735]$ GeV.
We let the distribution computed with the central replica 0 of the PDF set,
and with a fixed $\mwzero=80.385$ GeV value,
play the role of the experimental data ${\cal D}_{exp}$;
this choice does not spoil the validity of the method and of the conclusion and offers a sanity check on the fit results.
The covariance matrix is evaluated with the $N_{rep}$ replicas. We checked that the dependence
of the covariance matrix on $m_W$, in the interesting range of $\pm 20$ MeV around the central value of $80.385$ GeV;
is small and therefore we neglected it in the numerical analysis.
The statistical error on the pseudodata is estimated assuming 2 different luminosities,
$1$, and $300$ fb$^{-1}$.

Since the value of the PDF uncertainty affecting the $\mw$ determination
is sensitive to the fit window $[p_\perp^{min},p_\perp^{max}]$,
we perform a scan in the two values $p_\perp^{min,max}$ and plot, for each point in this plane,
the uncertainty value corresponding to the half-width of the $\Delta \chi^2 = 1$ interval.

\begin{figure}[t]
  \includegraphics[width=80mm,angle=0]{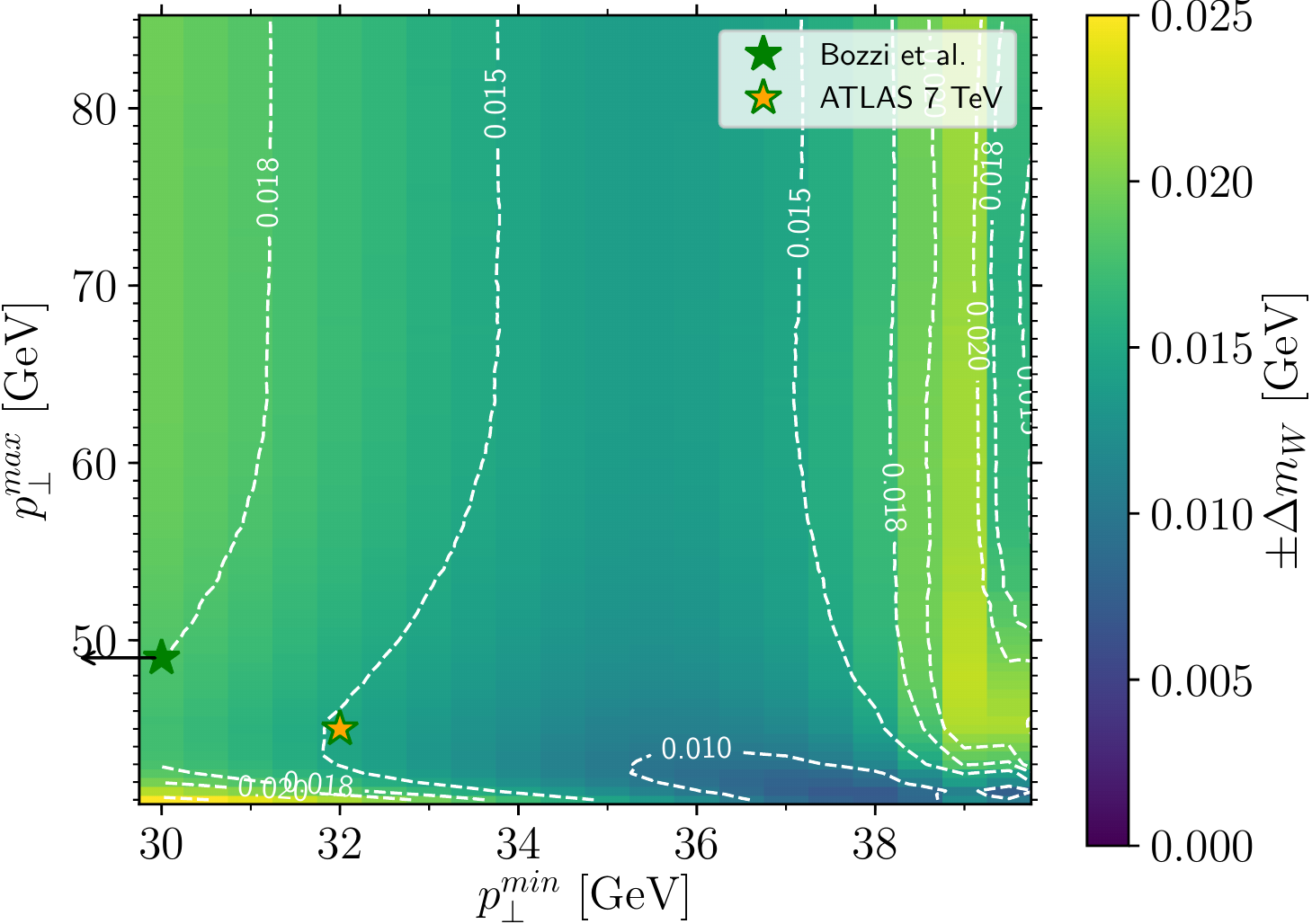}
  \caption{\label{fig:old-approach}
    PDF error estimated by computing the standard deviation of the $m_W$ values
    corresponding to the minima of the fit of $200$ replicas onto the template pseudodata represented
    by the central replica.}
\end{figure}

To present a comparison with the previous approaches,
we perform an analysis using the prescription of Eq.~\ref{eq:chi2nocov},
using 200 replicas, this time with a fixed $\mwzero=80.385$ GeV value, as distinct pseudodata distributions;
we generate the templates with the replica 0.
In Fig.~\ref{fig:old-approach} we show the analysis of distributions normalised
to the cross section integrated in the fitting interval.
The results, consistent with those presented in Ref.~\cite{Bozzi:2015hha} and labeled in all the Figures by "Bozzi et al.,
show a weak sensitivity to the upper limit of the fit window,
but  a clear dependence on its lower limit.

\begin{figure}[!t]
  \includegraphics[width=80mm,angle=0]{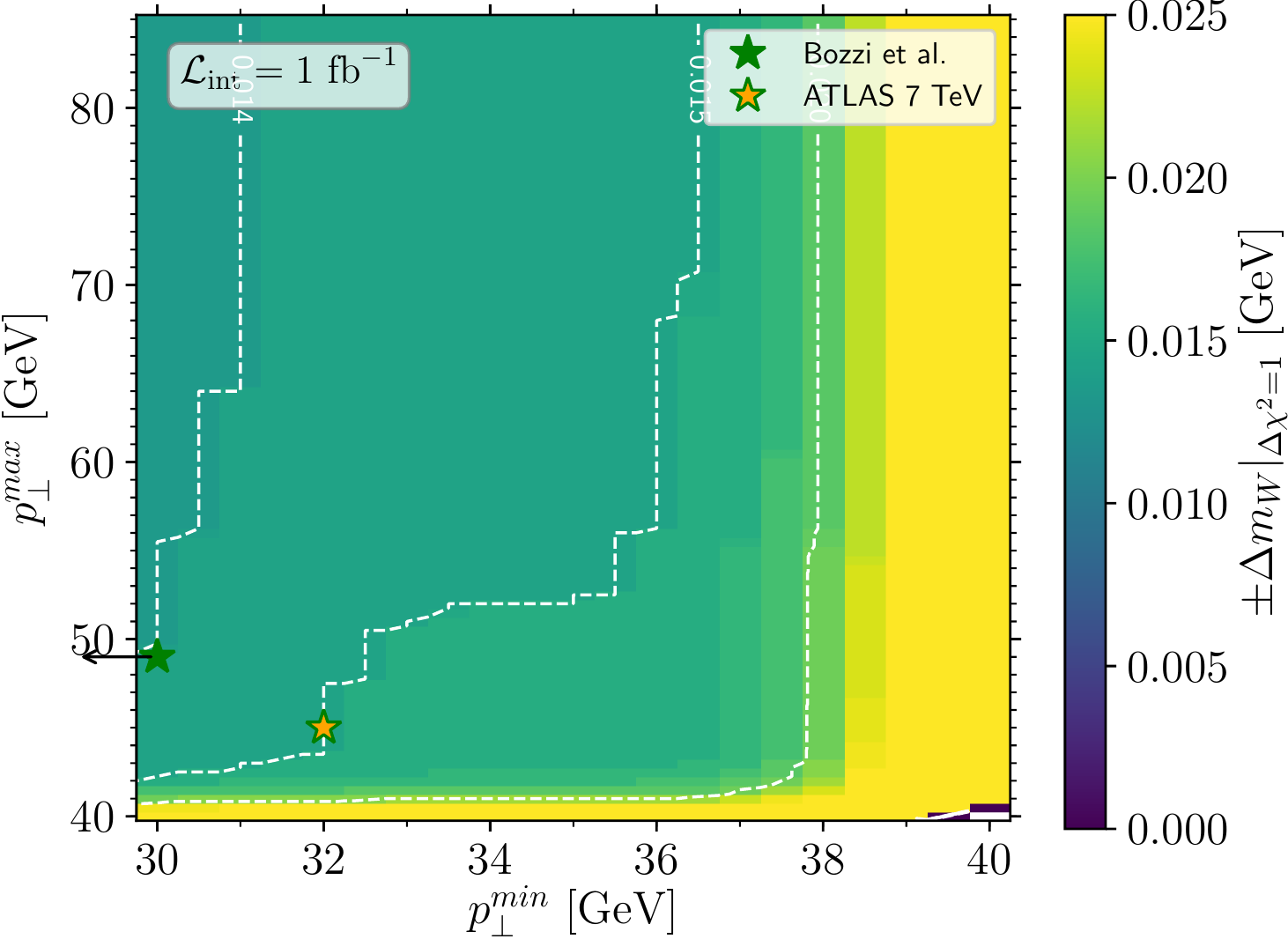}
  \caption{\label{fig:pdf-stat-1}
    PDF error as a function of the fit window expressed by its minimum and maximum $\ptlep$ values.
    Error estimated from a fit of shape distributions and using a covariance matrix obtained by summing
    the PDF one with a statistical (diagonal) error on the pseudodata corresponding to $1~\mathrm{fb}^{-1}$.
  }
\end{figure}

\begin{figure}[!t]
  \includegraphics[width=80mm,angle=0]{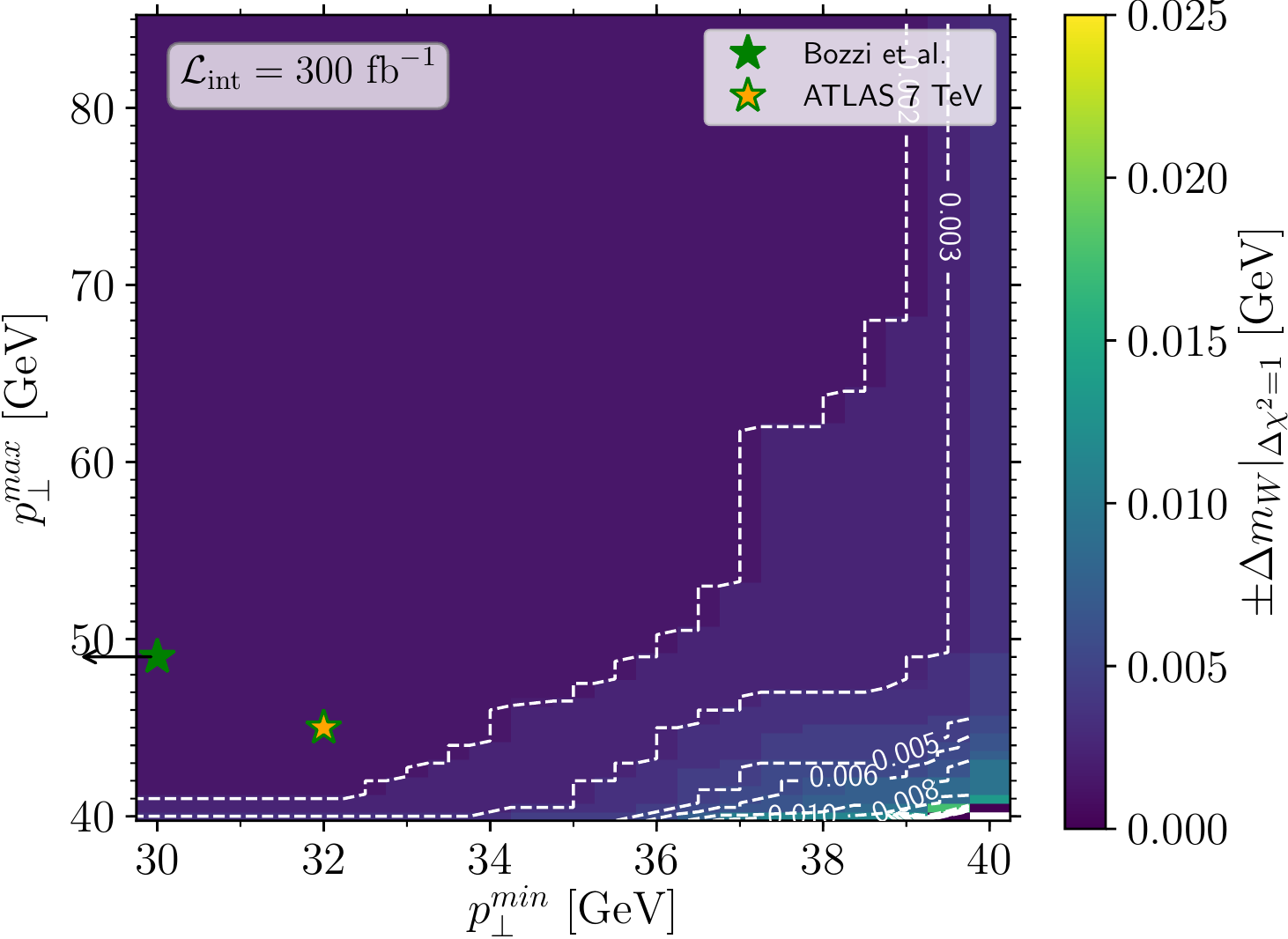}
  \caption{\label{fig:pdf-stat-300}
    Same as in Fig.~\ref{fig:pdf-stat-1}, but assuming $\mathcal{L}_{\mathrm{int}} = 300~\mathrm{fb}^{-1}$.
  }
\end{figure}

\begin{figure}[!t]
  \includegraphics[width=80mm,angle=0]{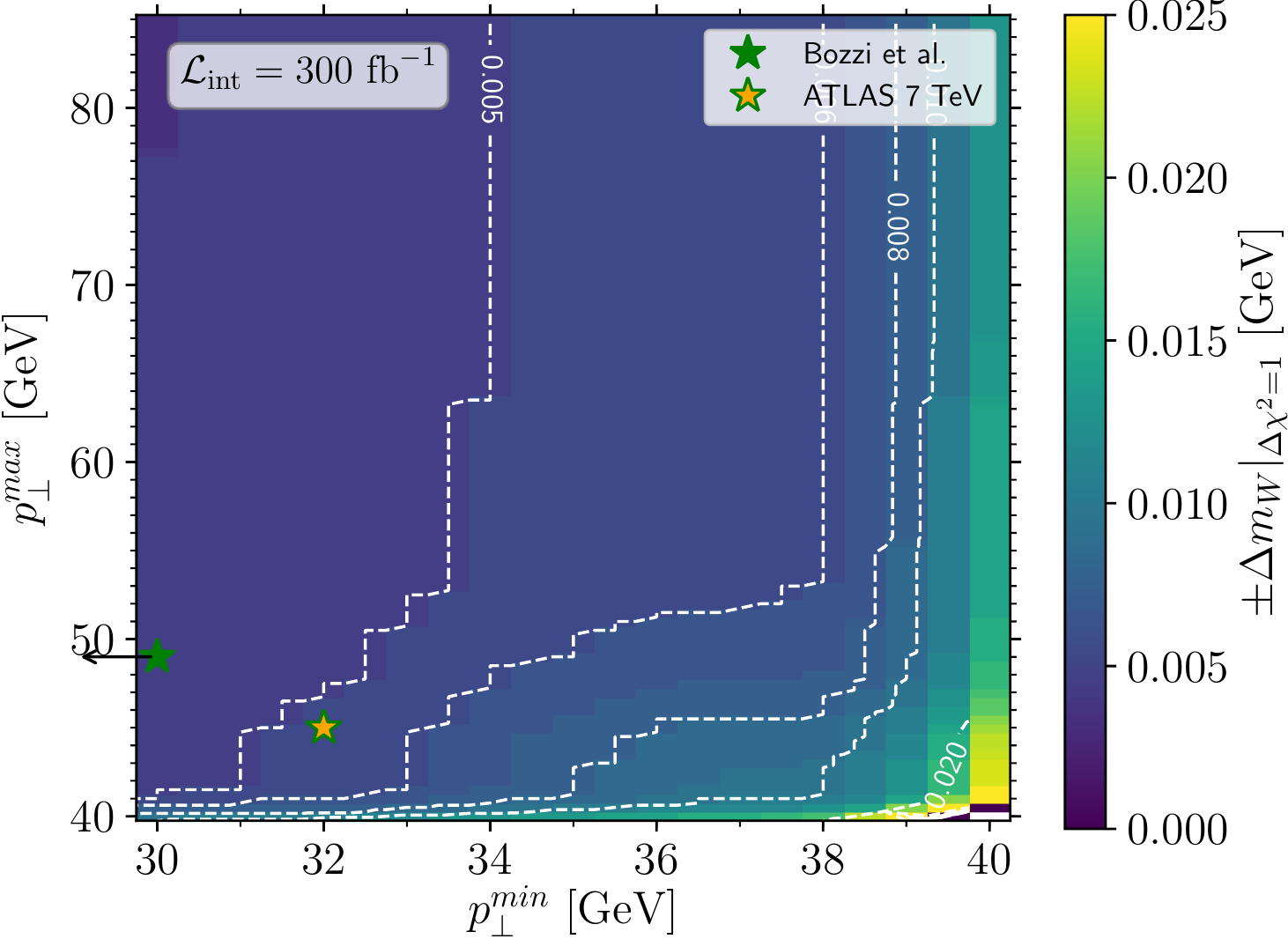}
  \caption{\label{fig:pdf-stat-300-mc}
    Same as in Fig.~\ref{fig:pdf-stat-300}, but including also a Monte Carlo error on the templates corresponding to $10^{10}$ events.
  }
\end{figure}

In Figs.~\ref{fig:pdf-stat-1},  \ref{fig:pdf-stat-300},
we present the results based on Eq.~\ref{eq:chi2cov},
in the case of normalised distributions,
assuming an experimental integrated luminosity $\mathcal{L}_{\mathrm{int}}$
respectively equal to $1$, and $300$ $\mathrm{fb}^{-1}$,
and no template Monte Carlo error.
Fig.~\ref{fig:pdf-stat-300-mc} also corresponds to $300~\mathrm{fb}^{-1}$,
but we now include a Monte Carlo error extrapolated to a statistics of $10^{10}$ events~\footnote{Our simulations
  were performed with only 15 millions events, due to our constrained computational resources. For the illustrative
  purpose of this note we rescaled the Monte Carlo error to the target statistics}.
The statistical error is dominant in Fig.~\ref{fig:pdf-stat-1}, while it becomes negligible at high luminosity,
putting in evidence a strong reduction of the PDF uncertainty, down to the ${\cal O}(1\, {\rm MeV})$ level.
The Monte Carlo error of the templates has a visible impact, as shown in Fig.~\ref{fig:pdf-stat-300-mc},
and would become negligible in a sample with 200 bilions events.
We remark the weak sensitivity of the results to the fit window.
\begin{figure}[!t]
  \includegraphics[width=80mm,angle=0]{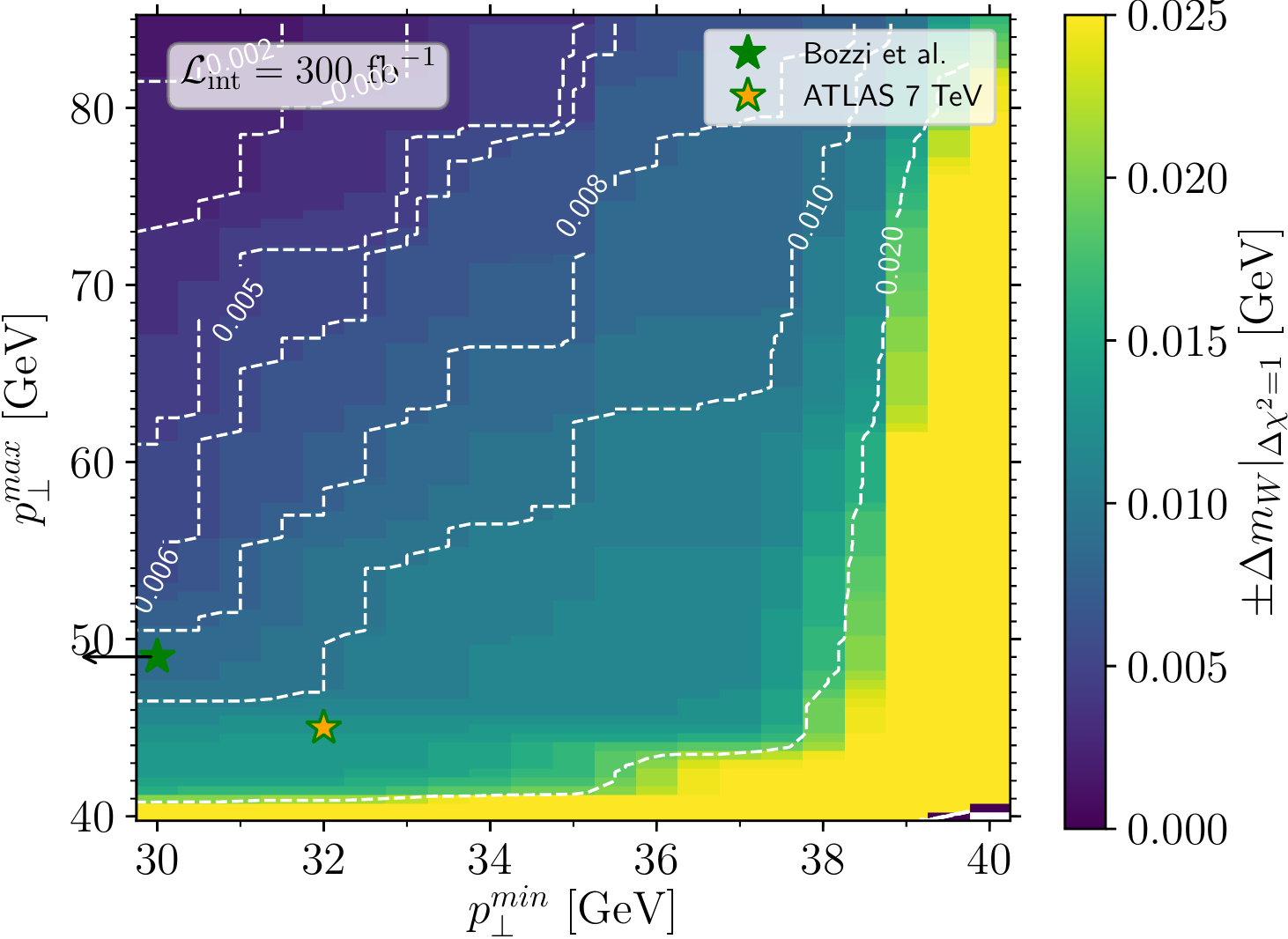}
  \caption{\label{fig:pdf-stat-300-expsyst}
    Same as in Fig.~\ref{fig:pdf-stat-300},
    but including also detector effects modelled according to Ref.~\cite{Manca:2016fpw}.
  }
\end{figure}
  We eventually show in Fig.~\ref{fig:pdf-stat-300-expsyst} the results
  corresponding to $300~\mathrm{fb}^{-1}$
  and a systematic error on the muon momentum reconstruction
  simulated via the model of Ref.~\cite{Manca:2016fpw}.
  The covariance matrix used in Fig.~\ref{fig:pdf-stat-300-mc}
  is added to the one coming from the detector effects,
  estimated using 100 toys.
  The negative impact in the description of the peak region balances the improved control
  on the tails of the distribution, increasing the size of the total error.
  We have checked that a reduction by a factor of $10$ of the gaussian smearing of the lepton momentum,
  would lead to uncertainties close to the ones shown in Fig.~\ref{fig:pdf-stat-300}.
Other sources of theoretical systematics,
such as perturbative QCD or parton shower uncertainties,
could become one of the limiting factor for the $\mw$ determination,
and will be considered in a future publication.

We observe that this approach strongly reduces the impact of PDF uncertainties
because of the specific structure of the bin-bin PDF covariance matrix $\Sigma_{PDF}$ of the $\ptlep$ distribution,
with the presence of quite distinct blocks formed by the bins below and above the Jacobian peak
\footnote{A large $N_{rep}$ value avoids the caveats of Ref.~\cite{Michael:1993yj}  }.
The eigenvalues spectrum of $\Sigma_{PDF}$ covers more than 7 orders of magnitude
between the largest and the smallest elements in absolute value.
The broad range of the eigenvalues induces a very narrow shape of the $\chi^2$ distribution as a function
of $\mw$, implying a strong penalty factor for all the templates that do not perfectly overlay
their peak position with the one of the data.
The penalty applies to the differences in the tails of the $\ptlep$ distribution,
while, at the same time, an excellent sensitivity to $\mw$,
at the 1 MeV level given by the templates granularity, is preserved,
as we explicitly verified as a sanity check of the approach.
The important role played by $\Sigma_{PDF}$ is partially smeared by the interplay between PDF and
statistical  and systematic errors.
Since $C=\Sigma_{PDF}+\Sigma_{stat}+\Sigma_{MC}+\Sigma_{exp,syst}$,
at low luminosities or low template accuracy the statistical error  has a non-trivial interplay
with the PDF error, yielding larger uncertainties than the values obtained for each class of errors alone;
at high-luminosities, with highly-accurate templates,
instead we approach the limit $C\simeq\Sigma_{PDF}$ and the corresponding strong uncertainty reduction.
Similar comments apply to the inclusion of the experimental systematic errors.

The PDF uncertainty band of the $\ptlep$ distribution is given
by a combination of perturbative and non-perturbative effects,
which can not be analytically separated;
although the pQCD elements (Dokshitzer-Gribov-Lipatov-Altarelli-Parisi equations, QCD sum rules) in the proton description
can not be qualified as uncertainty sources,
they nevertheless enter in the generation of the uncertainty band,
because of their entanglement with the data.
The covariance matrix allows the effective encoding of
a substantial piece of information of pQCD origin,
which should not be qualified as uncertainty, and includes it in the fit.
The description of the proton in terms of a QCD-inspired model
and the representation of the uncertainty via Monte Carlo replicas
are thus the two elements allowing the PDF uncertainty reduction.
The discussion of the PDF sets representing the associated uncertainty via Hessian eigenvectors
will be presented in a future publication.

In conclusion,
we have studied the theoretical systematic error
due to the PDF uncertainty,
focusing on the determination of the $W$ boson mass from the DY $d\sigma/dp_\perp^{\ell}$ distribution.
We included this systematics in the $\chi^2$ that we use in the data fitting,
achieving the automatic inclusion
of the bin-bin correlation with respect to PDF variations.
We observe a drastic reduction of the PDF uncertainty on $\mw$,
which we explain as a consequence of the strong kinematic correlation,
of pQCD origin,
of the bins above and below the Jacobian peak of the distribution.
The interplay of the PDF
with the statistical and experimental systematic errors
yields non trivial results, when the statistics is limited and systematics not fully understood.
We consider this approach promising in view of the reduction
of one of the bottlenecks limiting so far the high-precision determination
of $\mw$ at hadron colliders.
The formulation of Eq.~\ref{eq:chi2cov} is well suited
for a direct and efficient inclusion of the PDF uncertainty
in the analysis of the experimental data.
The use of this information should not be limited to the fit of $\mw$,
but it should also be part of the determination of any Lagrangian parameter derived in
the analysis of LHC observables.
The inclusion of further observables sensitive to the QCD model in a global fit, such as $\ptz$,
and of the corresponding cross-correlations, might provide additional
benefits to the \mw determination. However to properly assess the impact of such a procedure, a
detailed study beyond the scope of this article is needed.

\section*{Acknowledgements}
\begin{acknowledgments}
  E.B.~is supported by the Paul Scherrer Institut.
  We thank L.~Bianchini, S.~Camarda, E.~Manca, M.~Mangano, G.~Rolandi and L.~Silvestrini
  for useful discussions. We also thank the
  DESY IT department for making us available the computational resources of the BIRD/NAF2 cluster,
  which have been extensively used for this work.
  A.V.~is supported by the European Research Council
  under the European Unions Horizon 2020 research and innovation Programme (grant agreement number 740006).
\end{acknowledgments}

\bibliography{BV-PDF}

\end{document}